\documentclass[runningheads]{llncs}
\usepackage{graphicx}
\usepackage{caption}

\usepackage{pgfplots}
\pgfplotsset{compat=1.12}
\usepackage{multirow}
\begin{document}
\title{Partially Synthetic Data for Recommender Systems:
Prediction Performance and Preference Hiding}
%
%
\author{Manel Slokom\inst{1}\and 
Martha Larson\inst{1,2}\and
Alan Hanjalic\inst{1}
}
\authorrunning{M. Slokom et al.}
%
\institute{Delft University of Technology, Netherlands \and
Radboud University, Netherlands
\email{\{m.slokom,m.a.larson,a.hanjalic\}@tudelft.nl}}
\maketitle              
\begin{abstract}
This paper demonstrates the potential of statistical disclosure control for protecting the data used to train recommender systems. Specifically, we use a synthetic data generation approach to hide specific information in the user-item matrix. We apply a transformation to the original data that changes some values, but leaves others the same. The result is a partially synthetic data set that can be used for recommendation, but contains less specific information about individual user preferences. Synthetic data has a potential to be useful for companies, who are interested in releasing data to allow outside parties to develop new recommender algorithms, i.e., in the case of a recommender system challenge, and also reducing the risks associated with data misappropriation. Our experiments run a set of recommender system algorithms on our partially synthetic data sets as well as on the original data.
The results show that the relative performance of the algorithms on the partially synthetic data reflects the relative performance on the original data.
Further analysis demonstrates that properties of the original data are preserved under synthesis, but that for certain examples of attributes accessible in the original data are hidden in the synthesized data\footnote{This paper is accepted to Privacy in Statistical Databases 2020 (PSD) in the USB/INTRANET proceedings }.

\end{abstract}
\section{Introduction}
Since the privacy concerns raised by the Netflix Challenge~\cite{narayanan2008robust}, companies hesitate to release data to external researchers.
We investigate the potential of statistical disclosure control for the creation of synthetic data useful for recommender system research.
The paper elaborates on the idea of using synthetic data to evaluate recommender systems, which we previously introduced in~\cite{slokom2018comparing}.
The goal is to compare the relative performance of different recommender system algorithms without making use of the original data.
We investigate whether algorithms can be ranked in terms of their performance by training and testing on a partially synthesized version of the data.
Specifically, we apply a machine learning technique (CART) to create a partially synthesized data set that has two properties:
First, the relative performance of algorithms tested on the synthesized data reflects the relative performance of algorithms tested on the original data, and, second, attributes related to the preference of users that are accessible in the original data are hidden in the synthesized data.
The paper delivers a proof-of-concept that the use of synthetic data for recommender system research should not be ignored and is worth exploring in greater depth.

Our work is motivated by the observation that previous research on protecting user data in recommender systems has focused on specific threat models involving de-identification of anonymized data~\cite{sweeney2002achieving} or inference of sensitive information on protected data~\cite{weinsberg2012blurme}. In contrast, our work focuses on a new threat model: the data is neither anonymized nor otherwise protected. Instead, the goal of synthesis is to block the accessibility of specific attributes inherent in the user-item matrix.
Attribute blocking is desirable since it supports deniability, but also potentially lowers the incentives for attackers to misuse data. Such a threat model has recently become important, with the recent revelations by the mainstream media of the misuse of data acquired from Facebook~\cite{facebook}.

Statistical disclosure control is used in a variety of other fields in order to release data. The US Census Bureau (2006) has released a partially synthetic data for the survey of income and program participation by replacing quasi-identifiers for instances at high risk with imputation~\cite{abowd2001disclosure}. Also statistical agencies in Germany and New Zealand are developing synthetic data sets. We believe that the recommender system community has not yet looked at synthetic techniques for privacy protection because of its conventional focus on absolute levels of prediction performance.
The novel contribution of our work is to provide an empirical demonstration that the loss of absolute performance when data is partially synthesized is not necessarily great enough to render the data useless for recommendation.
The paper\footnote{Our code is available at: \url{https://github.com/SlokomManel/SynRec}} first positions our approach in its context (Section~\ref{sec:background}) and provides the necessary technical detail (Section~\ref{sec:experiments}).
Then, we discuss the analytical validity of the synthesized data (Section~\ref{sec:AV}), its ability to support the relative comparison of prediction performance (Section~\ref{sec:PP}), and also its ability to hide preference (Section~\ref{sec:DC}), and finish with a conclusion and an outlook.

\section{Background and Related Work}
\label{sec:background}

\subsection{Threat Model}
We specify our threat model, which, as mentioned above, differs from the conventional threat models addressed in recommender system research. 
The threat model is summarized in Table~\ref{tab:threat}, and its structure is inspired by the components specified in~\cite{Salter:1998}.
For precision we use this specific threat model, but also point to the more general attack models for recommender systems discussed by~\cite{burke2005identifying}. Put simply, the attacker has access to the user-item matrix containing ratings and knows the identity of the users and items.
The objective is to access information about users' attributes inherent in the matrix. We consider our synthetic data to be a success if we block access to this information.

\begin{table}[!ht]
\centering
\caption{Threat model addressed by our approach (see also~\cite{slokom2018comparing})}
\label{tab:threat}
\begin{tabular}{lllll}
\hline
\textbf{Component}       & \textbf{Description} \\ \hline
\textit{Adversary: Objective}  & Specific attributes of users inherent in the user-item matrix  \\ \hline
\textit{Adversary: Resources}  & Knowledge of items and users\\ \hline
\textit{Vulnerability:Opportunity} & Possession of clean-text user-item matrix  \\ \hline
\textit{Countermeasure} & Make access to original attributes unreliable \\ \hline
\end{tabular}
\end{table}

Our approach is intended to serve as an initial proof-of-concept. We designed it with full awareness of two limitations that future work will address. 
First, disclosure (i.e., when an adversary obtains previously unknown information about the target user) takes three forms, identity disclosure, attribute disclosure and inferential disclosure~\cite{templ2017statistical}. Here, we look only at specific examples of attribute disclosure. 
Second, the level that is necessary in practice is related to factors such as the exact definition of deniability, which we do not consider in this first proof-of-concept. Finally, note that we are addressing a use case in which only a single data set is synthesized. 
Multiple synthetic data sets would make possible a differential attack.

\subsection{Statistical Disclosure Control}
Statistical Disclosure Control techniques can be defined as the set of methods used to reduce the risk of disclosing information about users. 
Such methods are based on modifying or reducing data that is released.
Typical, micro data protection methods are classified into two categories based on how they manipulate the original data in order to build the protected data set~\cite{domingo2008survey}\cite{Torra2017}: First, masking methods, which generate a modified version of the original data, either with perturbations or other operations~\cite{shlomo2008protection}. 
Second, synthetic data generation (SDG) methods, which first construct a model of the data and then generate random artificial values from this model~\cite{domingo2008survey}. 
Several approaches have been proposed in the literature for generating synthetic data: data distortion by probability distribution~\cite{liew1985data} and synthetic data by multiple imputation~\cite{rubin1993discussion}. 
Recent techniques for synthetic data generation on statistical disclosure control can be divided into three basic categories~\cite{drechsler2011synthetic}\cite{drechsler2011empirical}\cite{patki2016synthetic}, namely partially synthetic methods, fully synthetic methods, and hybrid methods.

Here, we focus on a partially synthetic method, which allows us to stay close to the original data for our proof-of-concept. Fully synthetic methods could be explored in the future. We chose a machine learning method based on CART, because it is easy to implement, but also since it has been shown to perform well for census data used for demographic analysis~\cite{abowd2001disclosure}\cite{drechsler2011empirical}. CART is a decision tree based approach, and as such has the advantage of being a non-parametric method. The implication is that the dependencies to be learned in the data do not need to be specified in advance and it is possible to learn an increasing number of dependencies as the data set grows larger.

\subsection{Data Synthesis for Recommendation}
Previous use of synthetic data for recommender system research has been limited, and, to our knowledge, has never been explored for the purposes of protection before~\cite{slokom2018comparing}. 
The focus on synthetic data has been in research on context-aware recommendation. In~\cite{pasinato2013generating}, the authors proposed an abstract methodology for context-aware collection of data (in terms of item ratings and context of attributes). In~\cite{tso2006empirical}, the authors built a methodology to generate synthetic data sets for evaluating \textit{attribute-aware} recommender systems. However, they only focused on the generation of item attributes. In~\cite{del2017datagencars}, the authors designed a Java-based synthetic data set generator called \emph{DataGenCARS} to construct data sets for the evaluation of \textit{context-aware} recommendation systems, i.e., to complete the amount of context information characterizing the real ratings or to re-compute the ratings according to other specific user profiles. DataGenCARS was recently exploited by~\cite{del2017towards}.

\section{Generating Synthetic Data}
\label{sec:synthesis}

In this section, we explain how the CART method for data synthesis is applied to our recommender system data. 
A detailed formal description is available in~\cite{drechsler2011empirical}. 
Synthesis occurs in three steps: First, we designate ratings in our data set that will be retained in the synthetic data, and use these as training data. 
The training data is represented as a set of user-item pairs.
We train a tree by splitting the training data on the item ratings, each time optimizing the Gini Index. 
\begin{equation}
    Gini (A) = \sum _{i=1}^{C} p_{i}(1-p_{i})
\end{equation}

where $A$ is a node, $C$ is the number of classes in the node (i.e. Male/Female or ratings: 1--5), and $p_{i}$ is the class probability for the $i^{th} $ class.
Gini Index splits into groups to minimize the heterogeneity of values within groups. 
Finally, we allow the tree to classify each user-item pair for which the rating is to be generated.
The user-item pair is classified into a leaf, and the generated value is drawn from the rating values that occupy this leaf by using Bayesian bootstrap~\cite{rubin1981bayesian}. 
The method has two parameters which control the extent to which the synthesis is `partial'. 
In our case, this means the proportion of original ratings that are retained in the synthesized data. 
First, the order in which variables are synthesized. 
In our case this parameter has no impact, since we synthesize only one variable (the rating) for each user-item pair. 
Second, the stopping rules that dictate the number of observations (ratings) that are assigned to a node in the tree.

\section{Experimental Setup}
\label{sec:experiments}
In this section, we describe our data sets and different recommender system algorithms. 
\subsection{Data sets and Synthesis}
We test our models on two publicly available data sets. Statistics are summarized in Table~\ref{dataset}.

\begin{table}[!h]
\centering
\caption{Data sets }
\label{dataset}
\begin{tabular}{lllll}
\hline
\textbf{Data sets}       & \textbf{\#users} & \textbf{\#items} & \textbf{\#ratings} & \textbf{Density} \\ \hline
\textit{MovieLens 100k}  & 943     & 1.682   & 100.000   & 6.3\%   \\ \hline
\textit{GoodBook}        & 53.000  & 10.000  & 6.000.000 & 1.12\%  \\ \hline
\end{tabular}
\end{table}

We choose MovieLens 100k\footnote{https://grouplens.org/datasets/movielens/} because it is well-known and its properties are well understood by the research community.
We choose Goodbooks-10K\footnote{https://www.kaggle.com/philippsp/book-recommender-collaborative-filtering-shiny/data} as a larger, sparser data set from a different domain.

\subsection{Recommender Algorithms}
Recall that the goal of the recommender algorithms in this paper is not to demonstrate the absolute performance of the algorithms, but rather to evaluate if the relative performance of algorithms is the same on the original and on the synthesized data. For this purpose, we need a selection of classic recommender algorithms, ranging from baselines that are known not to yield state-of-the-art performance, to current algorithms. With this purpose in mind, we chose the following algorithms. For rating prediction, we choose \textit{KNN} and \textit{Centered KNN}, as well as two well-known rating prediction algorithms that are implemented in Surprise\footnote{http://surpriselib.com/} \textit{Slope One}~\cite{lemire2005slope} and \textit{Co-clustering}~\cite{george2005scalable}. For ranking prediction, we choose a well know algorithms commonly used to deal with collaborative filtering: matrix factorization \textit{(MF)} and Biased Matrix Factorization \textit{(BMF)}~\cite{Koren:2009} and Pairwise Ranking Factorization Machines
\textit{(BPRFM)}~\cite{Guo:2016:PRP}. The experiments are implemented using WrapRec~\cite{Loni:2014}. The three models were trained with (50,100,200,300) iterations and 20 latent factors. We used the Synthpop\footnote{https://cran.r-project.org/web/packages/synthpop/index.html} package for the generation of  synthetic data. The percentage of ratings retained from the original data set is $42 \%$ for the synthetic MovieLens data set and $29\%$ for the synthetic Goodbook data set.

\section{Analytical validity}
\label{sec:AV}
\begin{figure}[!h]
    \centering
    \begin{minipage}{0.45\textwidth}
        \centering
        \includegraphics[width=\textwidth]{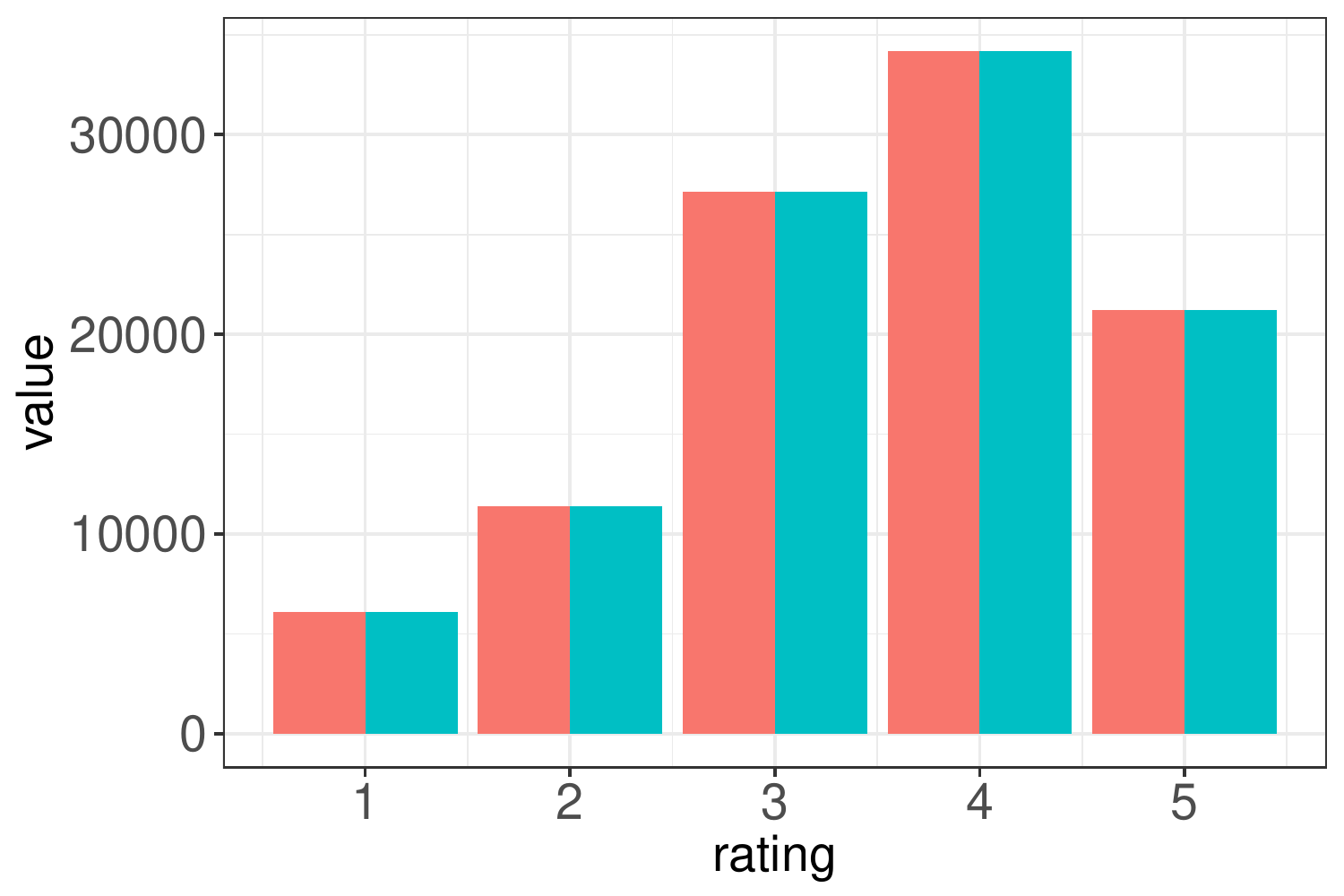} 
        \caption{Distribution of the percentage (original in the right bar VS synthetic in the left bar) of user ratings for MovieLens data set.}
        \label{fig:ratingdist}
    \end{minipage}\hfill
    \begin{minipage}{0.45\textwidth}
        \centering
        \includegraphics[width=\textwidth]{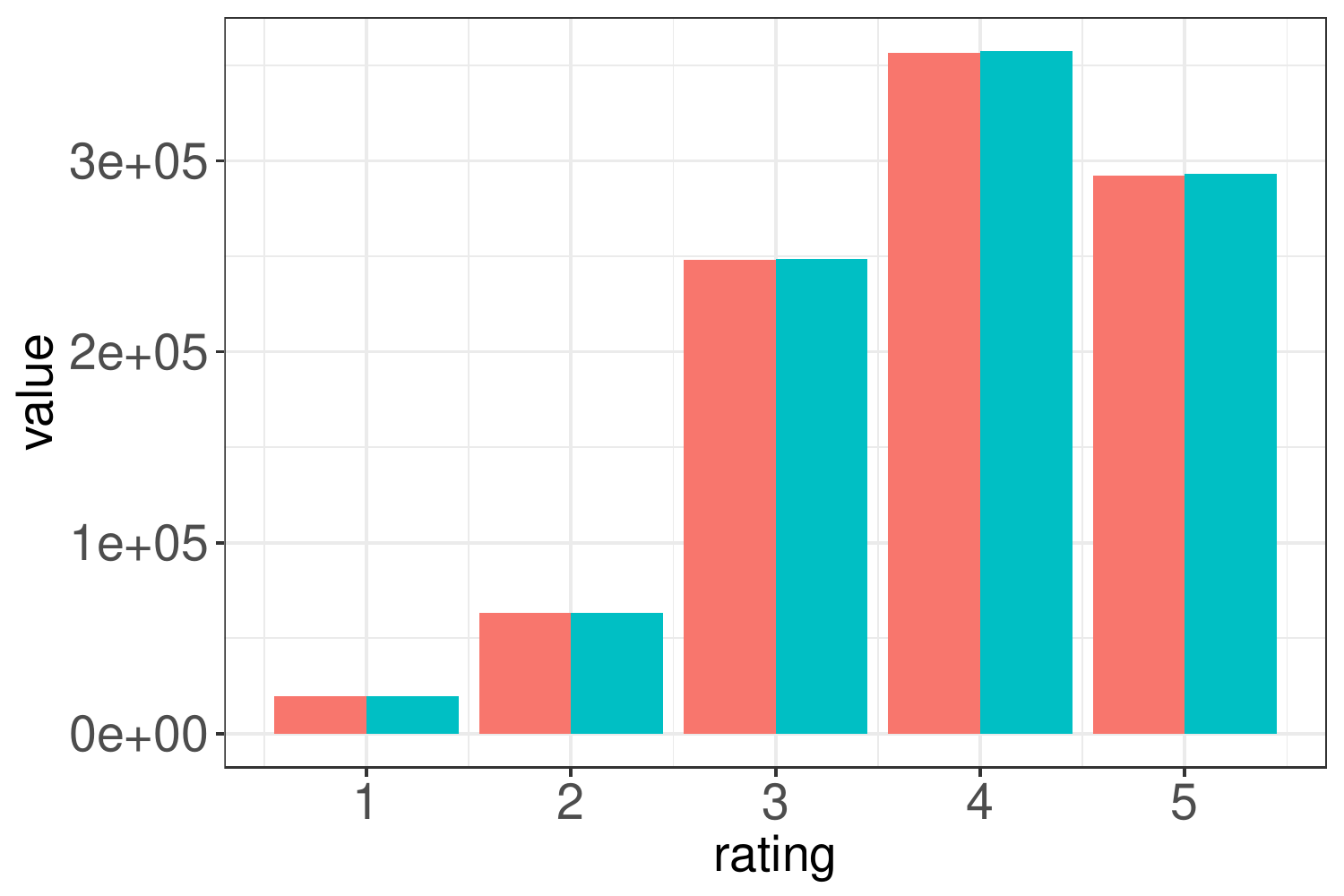} 
        \caption{Distribution of the percentage (original in the right bar VS synthetic in the left bar) of user ratings for GoodBook data set.}
    \end{minipage}
    \end{figure}

In the statistical disclosure control literature, the quality of synthetic data is measured by its \emph{analytical validity}, i.e., the degree of correspondence between global statistical properties of the original data and the synthesized data. In our work, the quality of the synthetic data will be measured by its ability to support the development of recommender system algorithms, and the degree to which user preference can be hidden.
However, before discussing these aspects, we provide statistics that reflect the analytical validity of the synthesized data in order to provide a picture of how the synthesis process is changing the data. In Figure~\ref{fig:ratingdist}, it can be seen that the distribution of rating values are nearly identical in the original and the synthesized data. Further, the top-10 most popular movies in MovieLens (the 10 items most frequently rated $>=4$), as well as the top-10 most popular directors and actors (whereby popularity was calculated on the basis of items rated $>=4$) and the most popular books in the Goodbook remain the same.

\section{Recommendation performance}
\label{sec:PP}
\subsection{Rating prediction}
\begin{figure}
    \centering
    \begin{minipage}{0.5\textwidth}
        \centering
        \includegraphics[width=\textwidth]{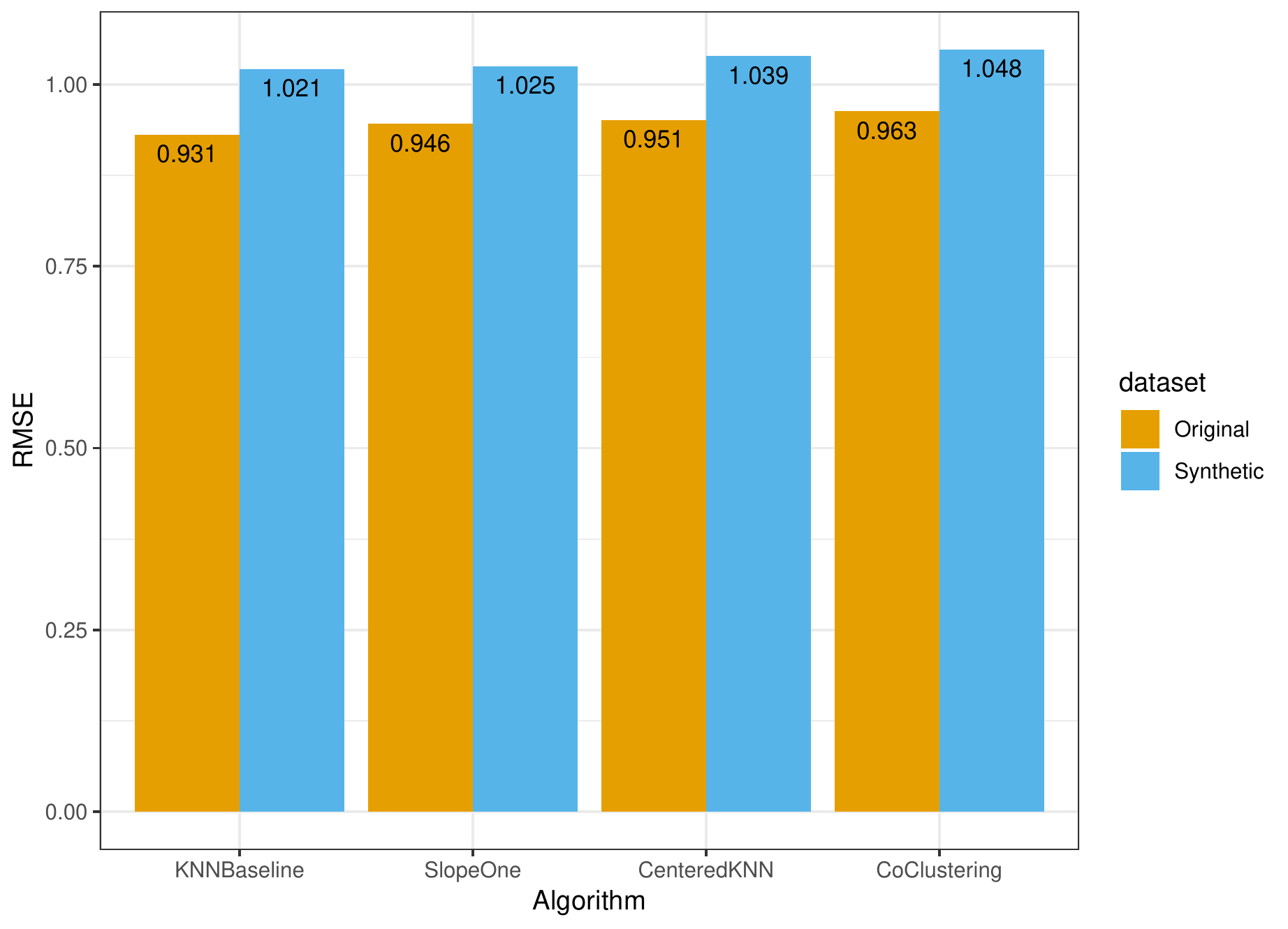} 
        \caption{RMSE for movieLens data set.}
        \label{fig:RMSEML}
    \end{minipage}\hfill
    \begin{minipage}{0.5\textwidth}
        \centering
        \includegraphics[width=\textwidth]{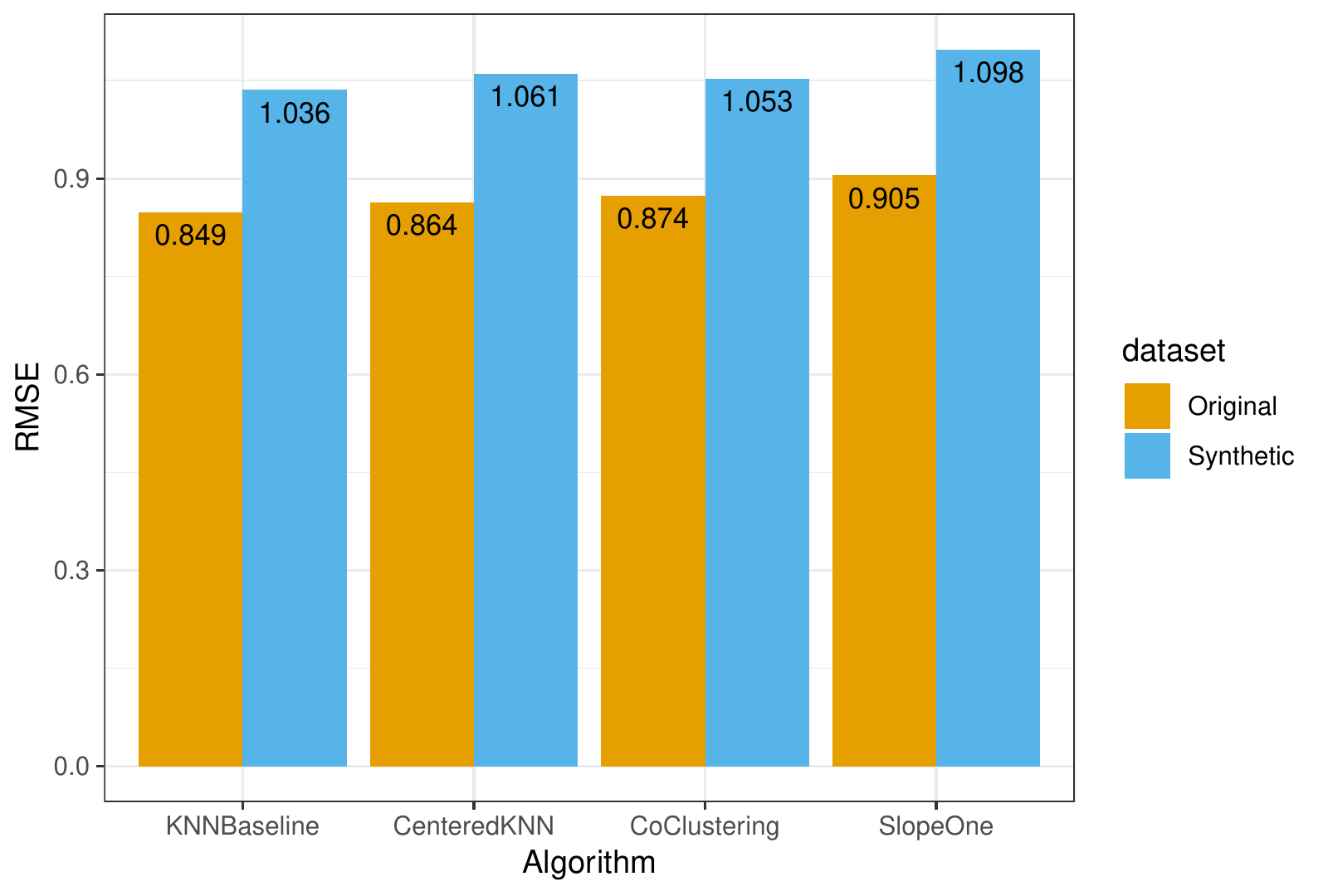} 
        \caption{RMSE for GoodBook data set.}
        \label{fig:RMSEGB}
    \end{minipage}
    \end{figure}

We start with classic rating prediction experiments in which we train and test algorithms on the original data set and compare them with algorithms that are trained and tested on the synthesized data. Results of our rating prediction algorithms for the MovieLens data set are given in Figure~\ref{fig:RMSEML}.

As expected, the absolute prediction performance, measured in terms of Root Mean Square Error (RMSE), on the synthetic data is lower than on the original data. However, the absolute performance is not directly of interest to us here. Rather we are interested to note that the relative performance of the four algorithms is the same, In other words, the co-clustering is the worst performing algorithm and KNN-Baseline is the best performing algorithm on both the synthesized and the original data sets. This result suggests that researchers can develop and test algorithms on synthetic data, and that an improvement of an algorithm on synthetic data will transfer to the original data. This result is quite striking in light of the known difficulty of transferring algorithms, for example, offline performance is known to be difficult to transfer to online settings. In the rest of the experiments, we test successively more challenging settings in order to understand the potential and limitations of this initial result.

In Figure~\ref{fig:RMSEGB}, the results of rating prediction on the GoodBook data set are shown.
Here, we again see that the best algorithm (KNNBaseline) and the worst algorithm (SlopeOne) is the same for both the original and the synthesized data. 
Note, however, that in this case, the order of the algorithms are not perfectly predicted. Coclustering and CenteredKNN are close in RMSE, and the fact that CeneteredKNN is better in the original data is not reflected in the synthesized data. This result suggests that additional investigation is necessary, but that all-in-all the relatively naive synthesis method used in this work holds promise for producing synthetic data useful for algorithm development. In the next section we turn to more challenging algorithms and more challenging task.

 \subsection{Ranking prediction} 
The results of the ranking prediction on the MovieLens and Goodbook Data Sets are reported in Tables~\ref{tab:rankmovies} and~\ref{tab:rankbook}. 

\begin{table}[!h]
\centering
\caption{Ranking performance for MovieLens Data set (Recall@5)}
\label{tab:rankmovies}
\begin{tabular}{clcccc}
\cline{3-6}
\multicolumn{1}{l}{}            &                    & \textit{NumIter=50} & \textit{NumIter=100} & \textit{NumIter=200} & \textit{NumIter=300} \\ \hline
\multirow{2}{*}{\textbf{MF}}    & \textit{Original}  & 0.0082              & 0.0022               & 0.0008               & 0.0005               \\ \cline{2-6} 
                                & \textit{Synthetic} & 0.003               & 0.0002               & 0.0                  & 0.0                  \\ \hline
\multirow{2}{*}{\textbf{BMF}}   & \textit{Original}  & 0.0181              & 0.0127               & 0.0103               & 0.0093               \\ \cline{2-6} 
                                & \textit{Synthetic} & 0.0137              & 0.0077               & 0.0058               & 0.0047               \\ \hline
\multirow{2}{*}{\textbf{BPRFM}} & \textit{Original}  & 0.068               & 0.0672               & 0.0682               & 0.0681               \\ \cline{2-6} 
                                & \textit{Synthetic} & 0.0664              & 0.0659               & 0.0678               & 0.067                \\ \hline
\end{tabular}
\end{table}


\begin{table}[!h]
\centering
\caption{Ranking performance for Goodbook Data set (Recall@5)}
\label{tab:rankbook}
\begin{tabular}{clcccc}
\cline{3-6}
\multicolumn{1}{l}{}            &                    & \textit{NumIter=50} & \textit{NumIter=100} & \textit{NumIter=200} & \textit{NumIter=300} \\ \hline
\multirow{2}{*}{\textbf{MF}}    & \textit{Original}  & 0.0003              & 0.0                  & 0.0                  & 0.0                  \\ \cline{2-6} 
                                & \textit{Synthetic} & 0.0001              & 0.0                  & 0.0                  & 0.0                  \\ \hline
\multirow{2}{*}{\textbf{BMF}}   & \textit{Original}  & 0.0049              & 0.005                & 0.005                & 0.0044               \\ \cline{2-6} 
                                & \textit{Synthetic} & 0.0047              & 0.0043               & 0.0041               & 0.0038               \\ \hline
\multirow{2}{*}{\textbf{BPRFM}} & \textit{Original}  & 0.0372              & 0.04                 & 0.0416               & 0.0427               \\ \cline{2-6} 
                                & \textit{Synthetic} & 0.0358              & 0.04                 & 0.0414               & 0.0415               \\ \hline
\end{tabular}
\end{table}


Again, a model was trained and tested on the original data and trained and tested on the synthesized data. Again, as expected, the absolute prediction performance, here measured as Recall@5, on the original data is better than on the synthesized data. What is interesting about these results is the relative performance of the recommender algorithms. For all cases, the best algorithm on the synthesized data is also the best algorithm on the original data, and the worst algorithm on the synthesized data is also the worst algorithm on the original data. Further, note that in every case, the original data and the synthetic data reflect each other as we search for the optimal number of iterations. The implication is that an optimum found on the synthesized data transfers to the original data. On the basis of these results we draw the conclusion that our proof-of-concept has successfully established the potential of partially synthetic data to support the comparison between recommender system algorithms, and is worthy of further, more detailed investigation.

\section{Disclosure control}
\label{sec:DC}
In this section, we look at the ability of synthesized data to hide the preferences of users. It is important to understand that we are not attempting to hide all preference information of a user. Because our goal is to have a synthetic data set that is useful for the purpose of recommendation, it would not make sense to remove all preference information from the data. Rather, we are interested in demonstrating that it is possible to hide specific information about individual users related to a particular aspect of preference. The overall goal is a demonstration that it is not necessary that a data set preserve all aspects of user preference in order to be useful for testing recommendation algorithms. Here, we first focus on two particular attributes that can be calculated from the user-item matrix and that reflect user preference: favorite actor and favorite director (for MovieLens) and favorite author (for Goodbook). Note that we are not asserting that these attributes are necessarily sensitive data. Instead, we chose these attributes as examples because of their accessibility (they are easy to calculate from the user-item matrix) and because of their plausibility (e.g., it is plausible that someone would want to hide information about their favorite actor or director in the wake of \#metoo revelations).

We choose a straightforward method for calculating the attributes for each user from the user-item matrix. For each user, we take all items that a user has rated $ >=4$ and consider these the ``preferred" items. For MovieLens, we access the IMDB website in order to determine the director and the actors that are associated with the movies. For GoodBook, the author information is included in the data set. We then calculate the most frequent director/actor/author on the set of the user's preferred items, and define this person to be the favorite director/actor/author for the user.

In order to examine disclosure control, we check the ability of the synthesized data to hide this preference information. For MovieLens, the percentage of users who had different favorite director in the original and the synthesized data is $64 \%$ and the percentage of users who had different favorite actor is $77 \%$. For Goodbooks, the percentage of users who had different favorite author is $72 \% $.

It is important to keep in mind that these results represent an initial demonstration that it is possible to hide preference information in the user-item matrix without the destroying the usefulness of the data for the purpose of developing recommender system algorithms. We do not claim that hiding favorite director/actor/author constitutes perfect disclosure control. Rather our point is that if a user would like to control this kind of preference information disclosure control is possible. Finally, we mention that the average absolute change in rating between the original data and the synthesized data is 0.825 for the MovieLens data set (variance 0.746) and 1.034 for the Goodbook data set (variance $0.814$). These statistics demonstrate that the synthetic data is quite far removed from the original data, at the level of the individual user. 

\section{Conclusion}
In this paper, we proposed a partially synthetic data generation technique that has two purposes: First it allows for the creation of data sets that can be used to develop and test algorithms. Our experimental results show that the relative performance of a set of recommender system algorithms developed on the synthetic data is reflected in their relative performance on the original data. Second, it makes it possible to hide certain preference-related user information otherwise accessible from the original user-item matrix.

We adopted the CART method for generating partially synthetic data because of its success in protecting the original data. However, we point out that we did not otherwise optimize it, e.g., to capture particular regularities in the data. We also treated ratings as categorical data and did not generate ratings for items not rated in the original data set. These points represent areas on which future work should concentrate. Our ultimate goal is a fully synthetic data set that represents enough of the characteristics of the original data to be used for recommender system research.

The idea and investigation presented here was originally inspired by the success of the ad hoc method to data protection that was used in the ACM RecSys Challenge 2016~\cite{Abel:2016}. We believe that research on data protection in recommender systems should have a systematic basis, and also be designed with respect to explicit threat models, such as the one that we investigate here.


%
%

%
\bibliographystyle{splncs04}
\bibliography{paper.bib}

\end{document}